# THE APPLICATION OF ARTIFICIAL NEURAL NETWORK FOR THE ASSESSMENT OF THERMAL PROPERTIES OF MULTILAYERED SEMICONDUCTOR STRUCTURE

*Zbigniew Suszyński [1], Mateusz Kosikowski [2], Radosław Duer [3]*

University of Technology Koszalin, Department of Electronics and Computer Science,
Śniadeckich 2, 75-453 Koszalin, Poland,

email: [1] zas@man.koszalin.pl , [2] kazuchiro@interia.pl, [3] rduer@man.koszalin.pl

**ABSTRACT**

This paper presents the method for determination of thermal properties of multi-layered structure basing upon experimental data acquired with the help of photo-acoustic technique in harmonic modulation mode. The method proposed is based upon the matching of theoretical complex contrast characteristic of the investigated object to the experimental contrast characteristic realised by artificial neural network

## 1. INTRODUCTION

The non-destructive investigations of thermal properties of solids are very important for the semiconductor industry. Especially, in the case of high power semiconductor devices, which are usually produced as multli-layered planar structures, the quality of adhesion between the layers is one of the most important factors of reliability of such a device. The areas of weak adhesion are characterised by high thermal impedance. Generally, thermal properties of weak adhesion area are highly diversified so it can be assumed that thermal non-uniformity is present there [1]. The assessment of thermal properties (such as diffusivity, effusivity, thickness of layer) of such area is related to resolving the inverse problem of deconvolution in order to estimate thermal impedance of non-uniformity area. That was the motivation for this contribution.

The method presented utilitizes Artificial Neural Network (ANN) to resolve the inverse problem. In this approach, well trained ANN identifies the set of thermal properties for which the experimental complex contrast characteristic (CTC) matches the best to the theoretical CTC characteristic obtained from the model.

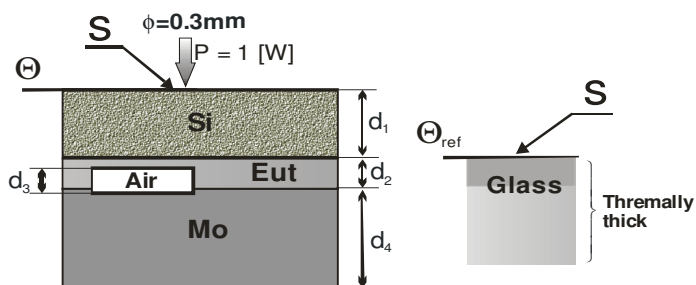

$d_1 \in \langle 0.3mm \; ; \; 0.4mm \rangle$

$d_2 \in \langle 0.04mm \; ; \; 0.1mm \rangle$

$d_1 + d_2 \in \langle 0.34mm \; ; \; 0.44mm \rangle$

$d_3 \in \langle 0.1 pm \; ; \; 5 \mu m \rangle$

$d_4 = 2mm$

*Fig.1. Profile of the experimental object.*

## 2. THE EXPERIMENT

### 2.1. The object of investigation

The object investigated during photo-acoustic experiment was typical high power thyristor structure. Three layers can be distinguished in such a structure: silicon-eutectics-molybdenum (Fig.1).

In the case of weak quality of adhesion the additional layer of high thermal impedance can be formed in the interface area between silicon and molybdenum. The thermally thick piece of passivated glass was used as a reference object for calculations of complex contrast characteristics.

## 2.2. The object of investigation

The experimental data were collected with the help of photo-acoustic technique in harmonic excitations mode with the standard measurement set-up [2]. The source of energy excitation was modulated laser beam of power about 1W. The frequencies of modulation were ranged from 106Hz to 586Hz. The signal of temperature response was acquired in both the amplitude and phase images [3]. The examples taken from the sequence of both the amplitude and phase images are presented in Fig.2. This figure also presents CTC images.

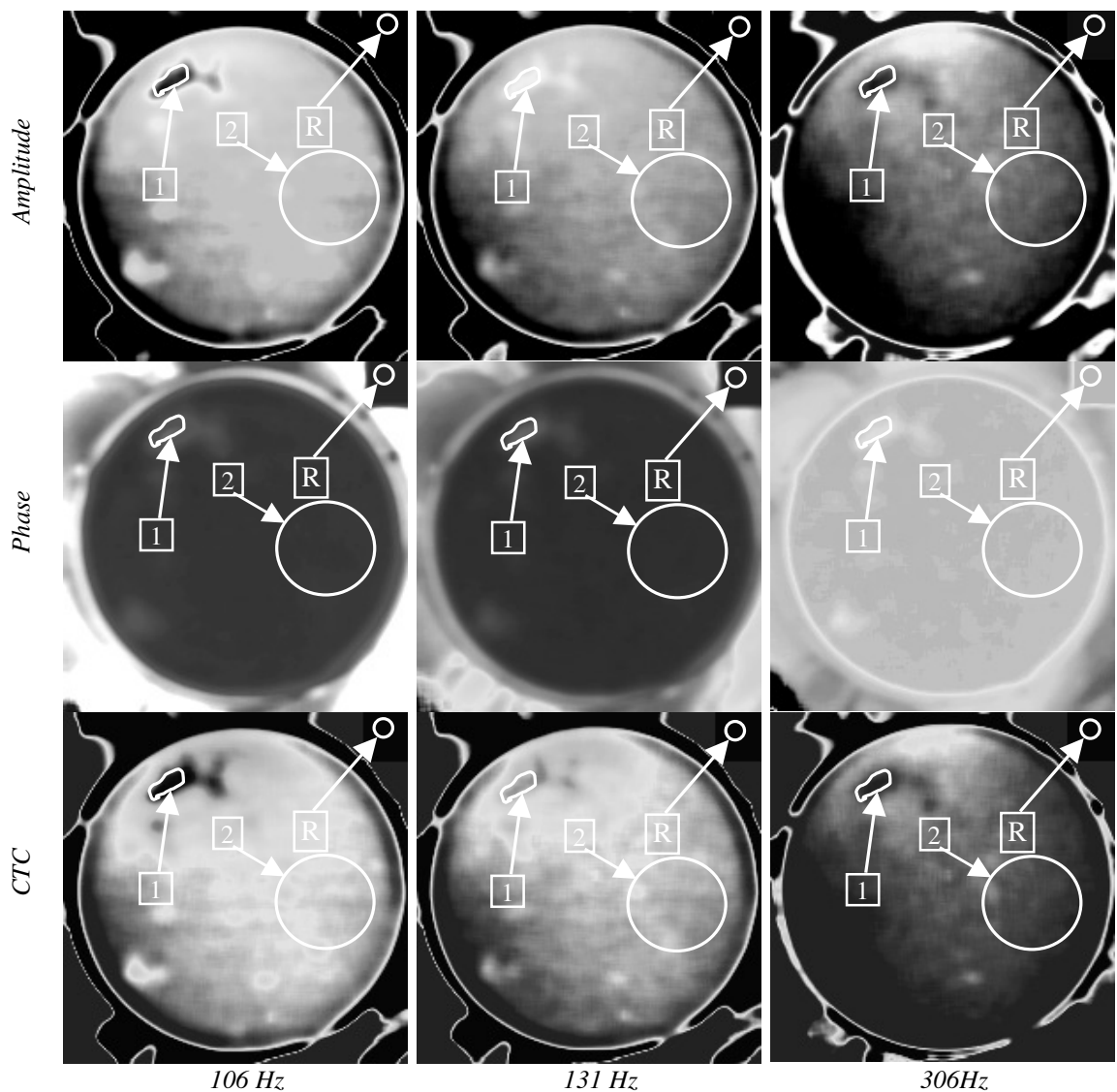

*106 Hz*   *131 Hz*   *306Hz*

*Fig.2. Examples of amplitude, phase and CTC images of investigated object*

For the purpose of proper analysis of thermal parameters of the object basing upon the results of photo-acoustic measurements it is mandatory to eliminate the influence of linear distortions introduced by each element of the photo-acoustic set-up. The best for that is the use of comparative method of analysis – for example Complex Thermal Contrast method [4, 5]. The mean value of temperature response signal from both the area of non-uniformity and the reference object were used further for determination of complex contrast characteristics.

## 2.3. 1-D TLM model

1-D TLM model was used in order to derive the theoretical complex contrast characteristics. The model utilities the electro-thermal analogy for determination of temperature response. In this case, the investigated structure is represented by four-terminal network (Fig. 3) [5, 6].

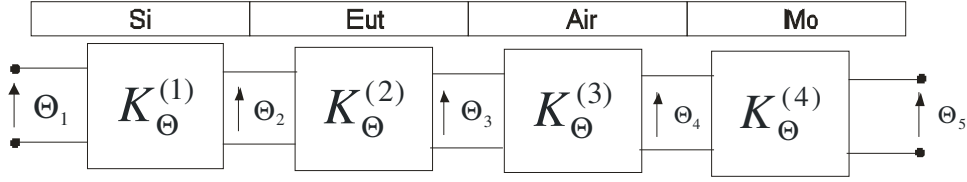

*Fig. 3. Four-terminal network model of 4-layer structure.*

$$\Theta_1 = \tilde{P} \cdot \left[ z_{11}^{(1)} - \cfrac{(z_{12}^{(1)})^2}{z_{11}^{(1)} + z_{11}^{(2)} - \cfrac{(z_{12}^{(2)})^2}{z_{11}^{(2)} + z_{11}^{(3)} - \cfrac{(z_{12}^{(3)})^2}{z_{11}^{(3)} + Z_C^{(4)}}}} \right] \quad (1)$$

$$z_{11} = Z_C \frac{\cosh(\Gamma)}{\sinh(\Gamma)} \qquad z_{12} = Z_C \frac{1}{\sinh(\Gamma)} \quad (2)$$

$$\Gamma = (1+i)\sqrt{\frac{\omega}{2\alpha}} \cdot d \qquad Z_C = (1+i)\frac{1}{S\varepsilon\sqrt{2\omega}} \quad (3)$$

$$\Theta_{n\,front} = \tilde{P} Z_C^{(1)} W_n \quad (4)$$

$$W_i = \begin{cases} \dfrac{\cosh \Gamma_1}{\sinh \Gamma_1} & \text{for } i=1 \text{ if thermally thin } (d_1 \leq \mu) \\ 1 & \text{for } i=1 \text{ if thermally thick } (d_1 \gg \mu) \\ \dfrac{\cosh \Gamma_{n-i+1}}{\sinh \Gamma_{n-i+1}} - \dfrac{\left(\dfrac{1}{\sinh \Gamma_{n-i+1}}\right)^2}{\dfrac{\cosh \Gamma_{n-i+1}}{\sinh \Gamma_{n-i+1}} + \dfrac{\alpha_{i-1}}{\varepsilon_{n+1,n-i+1}}} & \text{for } i \geq 2 \end{cases} \quad (5)$$

*Where:*

$\Theta_1$ - temperature of input of analyzed object [K], $\Gamma$ - operator of propagation, $Z_C$ - characteristic impedance [K/W], $\tilde{P}$ - stimulation power $\omega$ - frequency [Hz], $\varepsilon$ - thermal effusivity [$Ws^{0.5} \cdot m^{-2} \cdot K^{-1}$], $\alpha$ - thermal diffusivity [$m^2 \cdot s^{-1}$], $d$ - thickness of layer [m], n – number of layers, i – number of particular layer.

Temperature response at the surface of investigated object is given by the equation (1). It is possible to transform the equation (1) into more convenient form (4) using the recursive coefficient $W_i$ given by the equation (5) [7].

The values of temperature obtained were used for derivation of complex contrast characteristics.

Complex contrast is defined as difference between normalized complex temperatures for both the investigated and the reference objects. Thus, CTC presents the changes of both amplitude and phase of temperature disturbance [8].

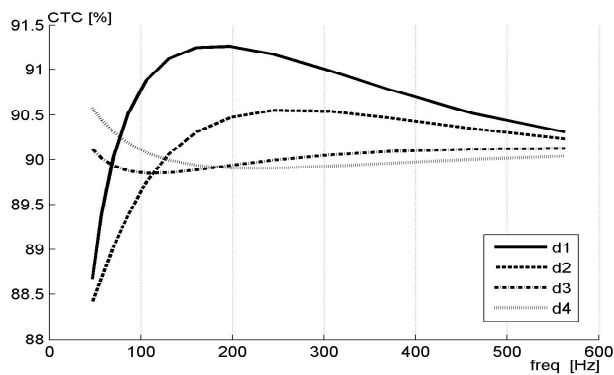
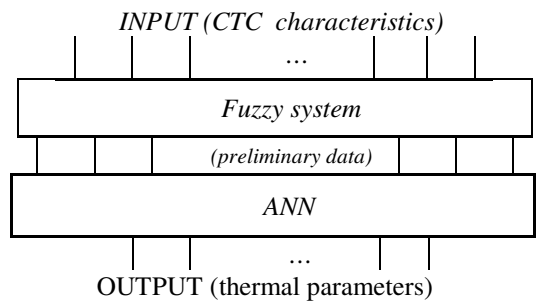

*a)* *b)*

*Fig. 4. Examples of CTC theoretical characteristics (a). Structure of designed system (b).*

|  |  | Area 1 (NNet result) | Area 1 (Real Value) | Area 2 (NNet result) | Area 2 (Real Value) |
|---|---|---|---|---|---|
| varuables | $d_{Si}$ | 319μm | *311μm* | 387 μm | 396μm |
| | $d_{Air}$ | 4.8 μm | 30.3 μm | *0.1 nm* | 0 |
| | $\varepsilon_{Eut}$ | 18068 | no data | 15087 | no data |
| | $\alpha_{Eut}$ | 8531 | no data | 8968 | no data |
| Const. | $\varepsilon_{Si}$ | 14000 | $\varepsilon_{Air}$ | 65 | $\varepsilon_{Mo}$ 18800 |
| | $\alpha_{Si}$ | 8e-5 | $\alpha_{Air}$ | 2.14e-5 | $\alpha_{Mo}$ 5.38e-5 |

*Tab. 1. The effusivity and diffusivity values of particular layer of investigated object appointed by ANN. The thickness ranges of each layer used for ANN training are shown in Fig.1.*

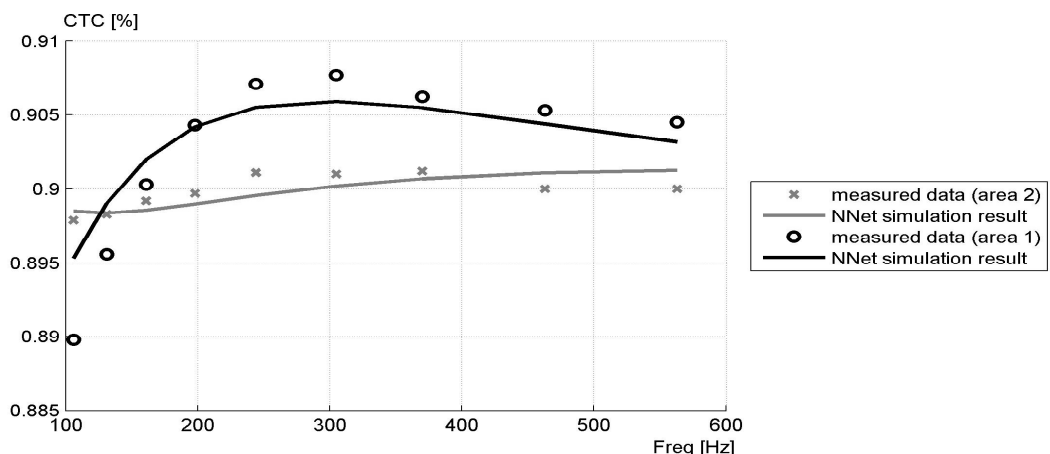

*Fig. 5. Example-results of pattern mach by ANN.*

## 3. THE RESULTS OF THE ANALYSIS OF THERMAL PROPERTIES

Fig.5 presents the comparison of CTC contrasts characteristics obtained from the ANN and the CTC contrasts characteristics values evaluated from experimental data. The relevant sets of geometrical and thermal parameters of particular layers of the object are presented in Tab.1. Solid line represents the CTC characteristics obtained from ANN and dots represents the values of CTC characteristics determined upon experimental data. The evaluated values of thickness of the silicon and the air layer are respectively: 319μm and 4.8μm (for the area of non-uniformity Fig.2 – area 1). These vales obtained for the uniform areas (Fig.2 – area 2) are as following: 387μm and 0.1nm.

For the purpose of verification of ANN simulation, the results of destructive investigations were used. Those investigations were performed for the same high power thyristor structure after photo-acoustic experiment. The results of this kind of investigations are presented in Tab.1 (Real Values). As it follows, the thickness of silicon layer as well as the thickness of air layer is in high convergence for the uniform area. The zero value means that spacial resolution of destructive investigation method was to low so it could not be compared with the value of 0.1nm achieved for the ANN method

Concerning the area of non-uniformity, both of the geometrical parameters investigated are different from each other. For the case of non-uniformity area it is caused by the insensibility of the model for the air layer thickness above 5μm so it was the biggest thickness of air layer assumed during thermal modelling and training of ANN.

## 4. CONCLUSIONS

The final results obtained are encouraging concerning the fact that the convergence of experimental characteristics to the ones obtained from ANN was good. Differences can be caused by the use of 1D thermal model. The limitations of 1D model in comparison to the 2D model can be significant. The solution for that can be usage of 2D or 3D thermal models for the purpose of ANN training. However, it will be related with extensive increase of the independent variables as well as the consumption of computer system resources. Nevertheless, the method for thermal properties analysis presented allows to utilise the advantages of CTC contrast method and the quickness of artificial neural networks. The solution of inverse problem given that way is insensible to the disturbances of measurement data as well as measurement errors. The trained ANN produces relevant and precise enough results even with significantly wrong value of experimental CTC characteristics for some frequency, which can appear due to large error made during the measurements for this frequency.